\documentclass[usenatbib]{mn2e}

\usepackage{amsmath, amssymb}
\usepackage{array}
\usepackage{graphicx}

\begin{document}

\newcommand{\be}{\begin{equation}}
\newcommand{\ee}{\end{equation}}
\newcommand{\pdf}{p}
\newcommand{\data}{{d}}
\newcommand{\Df}{{D}}
\newcommand{\Bf}{\mathcal{B}_{01}}
\newcommand{\lnBf}{\ln\Bf}
\newcommand{\mdl}{{M}}
\newcommand{\lsim}{\,\raise 0.4ex\hbox{$<$}\kern -0.8em\lower 0.62ex\hbox{$\sim$}\,}
\newcommand{\gsim}{\,\raise 0.4ex\hbox{$>$}\kern -0.7em\lower 0.62ex\hbox{$\sim$}\,}

\newcommand{\params}{{\theta}}
\newcommand{\mean}{{\mu}}
\newcommand{\like}{L}
\newcommand{\lnlike}{\mathcal{L}}
\newcommand{\ML}{^*}
\newcommand{\dr}{\textrm{d}}
\newcommand{\ie}{i.e.}
\newcommand{\reion}{\text{re}}

\newcommand{\cd}{\cdot}
\newcommand{\cds}{\cdots}
\newcommand{\ip}{\int_0^{2\pi}}
\newcommand{\al}{\alpha}
\newcommand{\ba}{\beta}
\newcommand{\de}{\delta}
\newcommand{\De}{\Delta}
\newcommand{\ep}{\epsilon}
\newcommand{\Ga}{\Gamma}
\newcommand{\ka}{\tau}
\newcommand{\io}{\iota}
\newcommand{\La}{\Lambda}
\newcommand{\Om}{\Omega}
\newcommand{\om}{\omega}
\newcommand{\si}{\sigma}
\newcommand{\Si}{\Sigma}
\newcommand{\te}{\theta}
\newcommand{\ze}{\zeta}
\newcommand{\vth}{\ensuremath{\vartheta}}
\newcommand{\vph}{\ensuremath{\varphi}}
\newcommand{\MM}{\mbox{$\cal M$}}
\newcommand{\tr}{\mbox{tr}}
\newcommand{\hor}{\mbox{hor}}
\newcommand{\grad}{\mbox{grad}}
\newcommand{\cx}{\ensuremath{\mathbf{\nabla}}}
\newcommand{\lap}{\triangle}
\newcommand{\arctg}{\mbox{arctg}}
\newcommand{\bm}[1]{\mbox{\boldmath $#1$}}
\newcommand{\eff}{{\rm eff}}
\newcommand{\tto}{\Rightarrow}
\newcommand{\lag}{\langle}
\newcommand{\rag}{\rangle}
\newcommand{\fiso}{f_{\text{iso}}}
\newcommand{\Afiso}{\vert f_{\text{iso}}\vert}
\newcommand{\norm}[3]{{N_{#1,#2}(#3)}}
\newcommand{\fexp}{\boldsymbol{e}}
\newcommand{\KL}{D_{KL}}
\newcommand{\ns}{n_S}
\newcommand{\omk}{\Om_\kappa}
\newcommand{\eps}{\epsilon}
\title[Applications of Bayesian model selection]{Applications of Bayesian model selection to cosmological parameters}

\author[Roberto Trotta]{Roberto Trotta\thanks{E-mail address: {\tt rxt@astro.ox.ac.uk}}
\\
Oxford University, Astrophysics,  Denys Wilkinson
Building, Keble Road, OX1 3RH, United Kingdom\\
D\'epartement de Physique Th\'eorique, Universit\'e de Gen\`eve,
24 quai Ernest Ansermet, 1211 Gen\`eve 4, Switzerland}

\maketitle

\begin{abstract}
Bayesian model selection is a tool to decide whether the
introduction of a new parameter is warranted by data. I argue that
the usual sampling statistic significance tests for a null
hypothesis can be misleading, since they do not take into account
the information gained through the data, when updating the prior
distribution to the posterior. On the contrary, Bayesian model
selection offers a quantitative implementation of Occam's razor.

I introduce the Savage--Dickey density ratio, a computationally
quick method to determine the Bayes factor of two nested models
and hence perform model selection. As an illustration, I consider
three key parameters for our understanding of the cosmological
concordance model. By using WMAP 3--year data complemented by
other cosmological measurements, I show that a non--scale
invariant spectral index of perturbations is favoured for any
sensible choice of prior. It is also found that a flat Universe is
favoured with odds of $29:1$ over non--flat models, and that there
is strong evidence against a CDM isocurvature component to the
initial conditions which is totally (anti)correlated with the
adiabatic mode (odds of about $2000:1$), but that this is strongly
dependent on the prior adopted.

These results are contrasted with the analysis of WMAP 1--year
data, which were not informative enough to allow a conclusion as
to the status of the spectral index. In a companion paper, a new
technique to forecast the Bayes factor of a future observation is
presented.

\end{abstract}


\begin{keywords}
Cosmology -- Bayesian model comparison -- Statistical methods --
Spectral index -- Flatness -- Isocurvature modes
\end{keywords}

\section{Introduction}

In the epoch of precision cosmology, we often face the problem of
deciding whether or not cosmological data support the introduction
of a new quantity in our model. For instance, we might ask whether
it is necessary to consider a running of the spectral index, an
extra isocurvature mode, or a non-constant dark energy equation of
state. The status of such additional parameters is uncertain, as
often sampling (frequentist) statistics significance tests do not
allow them to be ruled out with high confidence. There is a large
body of work\footnote{A good starting point is the collection of
references available from the website of David R.~Anderson,
Department of Fishery and Wildlife Biology, Colorado State
University. \label{priorrefs}} that addresses the difficulties
arising from the use of p--values (significance level) in
assessing the need for a new parameter. Many weaknesses of
significance tests are clarified, and some even overcome, by
adopting a Bayesian approach to testing. In this work, we take the
viewpoint of Bayesian model selection to determine whether a
parameter is needed in the light of the data at hand.

The key quantity for Bayesian model comparison is the marginal
likelihood, or evidence, whose calculation and interpretation is
attracting increasing attention in cosmology and
astrophysics~\citep{Drell:1999dx,Saini:2003wq,Lazarides:2004we,Beltran:2005xd,Kunz:2006mc,Trotta:2006pw},
after it was introduced in the cosmological context
by~\cite{Jaffe:1995qu,Slosar:2002dc}. The marginal likelihood has
proved useful in other contexts, as well, for instance consistency
checks between data sets~\citep{Hobson:2002zf,Marshall:2004zd},
the detection of galaxy clusters via the Sunayev-Zel'dovich
effect~\citep{Hobson:2002de} and neutrino emissions from type II
supernovae~\citep{Loredo:2001rx}. In this paper we use the
Savage--Dickey density ratio for an efficient computation of
marginal likelihoods ratios (Bayes factor), while in a companion
paper \citep{Trottaprep} we present a new method to forecast the
Bayes factor probability distribution of a future observation,
called PPOD (for ``Predictive Posterior Odds
Distribution'')\footnote{The method was called ExPO for ``Expected
Posterior Odds'' in a previous version of this
work~\citep{Trotta:2005ar}. I am grateful to Tom Loredo for
suggesting the new, more appropriate name.}. We then illustrate
applications to some important parameters of current cosmological
model building.

This paper is organized as follows: we review the basics of
Bayesian  model comparison in section \ref{sec:Bayesian} and we
introduce the Savage--Dickey density ratio (SDDR) for the
computation of the Bayes factor between two nested models. Section
\ref{sec:application} is devoted to the application of model
selection to three central parameters of the cosmological
concordance model: the spectral tilt of scalar perturbations, the
spatial curvature of the Universe and a totally (anti)correlated
isocurvature CDM contribution to the initial conditions. We
discuss our results and summarize our conclusions in section
\ref{sec:conclusions}.

Some complementary material is presented in the appendices. An
explicit illustration of Lindley's paradox is given in
appendix~\ref{app:lindley}, the mathematical derivation of the
SDDR is presented in appendix~\ref{app:sddr} while a series of
benchmark tests for the accuracy of the SDDR are carried out in
appendix~\ref{app:accuracy}.

\section{Bayesian model comparison}
\label{sec:Bayesian}

In this section, we first briefly review the basics of Bayesian
inference and model comparison and introduce our notation. We then
present the Savage--Dickey density ratio for a quick computation
of the Bayes factor of two nested models.

\subsection{Bayes factor}

Bayesian inference (see e.g.~\cite{JaynesBook,MKbook}) is based on
Bayes' theorem, which is a consequence of the product rule of
probability theory:
 \be \label{eq:Bayes_Theorem}
    p(\params | \data,\mdl)= \frac{p(\data | \params,\mdl)
    \pi(\params \vert \mdl)}{p(\data | \mdl)}.
\ee On the left-hand side, the posterior probability for the
parameters $\params$ given the data $\data$ under a model $\mdl$
is proportional to the likelihood $p(\data | \params,\mdl)$ times
the prior probability distribution function (pdf), $\pi(\params |
\mdl)$, which encodes our state of knowledge before seeing the
data. In the context of model comparison it is more useful to
think of $\pi(\params | \mdl$) as an integral part of the model
specification, defining the prior available parameter space under
the model $\mdl$. The normalization constant in the denominator of
\eqref{eq:Bayes_Theorem} is the {\em marginal likelihood for the
model $\mdl$} (sometimes also called the ``evidence'') given by
 \be \label{eq:evidence_def}
 p(\data | \mdl) = {\int_{\Omega} p(\data | \params, \mdl)
\pi(\params | \mdl)\dr\params}
 \ee
where $\Omega$ designates the parameter space under model $\mdl$.
In general, $\params$ denotes a multi--dimensional vector of
parameters and $\data$ a collection of measurements (data
covariance matrix, etc), but to avoid cluttering the notation we
will stick to the simple symbols introduced above.

Consider two competing models $\mdl_0$ and $\mdl_1$ and ask what
is the posterior probability of each model given the data $\data$.
By Bayes' theorem we have
\begin{equation}
\pdf(\mdl_i\vert\data)\propto
\pdf(\data\vert\mdl_i)\pi(\mdl_i)~~(i=0,1),
\end{equation}
where $\pdf(\data\vert\mdl_i)$ is the marginal likelihood for
$\mdl_i$ and $\pi(\mdl_i)$ is the prior probability of the $i$th
model before we see the data. The ratio of the likelihoods for the
two competing models is called the {\it Bayes factor}:
\begin{equation} \label{eq:bfac}
B_{01}\equiv\frac{\pdf(\data\vert\mdl_0)}
{\pdf(\data\vert\mdl_1)},
\end{equation}
which is the same as the ratio of the posterior probabilities of
the two models in the usual case when the prior is presumed to be
noncommittal about the alternatives and therefore
$\pi(\mdl_0)=\pi(\mdl_1)=1/2$. The Bayes factor can be interpreted
as an automatic Occam's razor, which disfavors complex models
involving many parameters (see e.g.~\cite{MKbook} for details). A
Bayes factor $B_{01} > 1$ favors model $\mdl_0$ and in terms of
betting odds it would prefer $\mdl_0$ over $\mdl_1$ with odds of
$B_{01}$ against 1. The reverse is true for $B_{01} < 1$.

It is usual to consider the logarithm of the Bayes factor, for
which the so--called ``Jeffreys' scale'' gives empirically
calibrated levels of significance for the strength of
evidence~\citep{Jeffreys,Kass}, $\vert \ln B_{01} \vert > 1;
> 2.5; > 5.0$. Different authors use different conventions
to qualify the Jeffreys' levels of strength of evidence. In this
work we will use the convention summarized in Table~\ref{Tab:Jeff}
-- often in the literature one deems odds above $|\ln B_{01}| = 5$
to be `decisive', but we prefer to avoid the use of the term
because of the strong connotation of finality that it carries with
it. If we assume that the two competing models are exhaustive,
\ie~that $p(\mdl_0|\data) + p(\mdl_1|\data) = 1$ and a
non--committal prior $\pi(\mdl_0) = \pi(\mdl_1) = 1/2$, we can
relate the strength of evidence to the posterior probability of
the models,
 \be
 \begin{aligned}
 p(\mdl_0|\data) & = \frac{B_{01}}{B_{01} + 1} \\
 p(\mdl_1|\data) & = \frac{1}{B_{01} + 1}.
 \end{aligned}
 \ee
 This probability is indicated in the third column of
Table~\ref{Tab:Jeff}.

The subject of hypothesis testing has received an enormous amount
of attention in the past, and the controversy on the subject is
far from being resolved among statisticians. An illustration of
the difference between Bayesian model selection and frequentist
hypothesis testing is given in Appendix~\ref{app:lindley}, where
Lindley's paradox is worked out with the help of a simple example.
There it is shown that the Bayesian approach has the advantage of
taking into account the information provided by the data, which is
ignored by frequentist hypothesis testing.
\begin{table}
\caption{Jeffreys' scale for the strength of evidence when
comparing two models, $\mdl_0$ versus $\mdl_1$, with our
convention for denoting the different levels of evidence. The
probability column refers to the posterior probability of the
favoured model, assuming non--committal priors on the two
competing models, \ie~$\pi(\mdl_0) = \pi(\mdl_1) = 1/2$ and that
the two models exhaust the model space, $p(\mdl_0|\data) +
p(\mdl_1|\data) = 1$.\label{Tab:Jeff} }
\begin{tabular}{l l l l}
$|\ln B_{01}|$ & Odds & Probability & Notes \\\hline
 $<1.0$ & $\lsim 3:1$ & $<0.750$ & Inconclusive \\
 $1.0$ & $\sim 3:1$ & $0.750$ & Positive evidence \\
 $2.5$ & $\sim 12:1$ & $0.923$ & Moderate evidence \\
 $5.0$ & $\sim 150:1$ & $0.993$ & Strong evidence \\
\end{tabular}
\end{table}

Evaluating the marginal likelihood integral of
Eq.~\eqref{eq:evidence_def} is in general a computationally
demanding task for multi--dimensional parameter spaces.
Thermodynamic integration is often the method of choice, whose
computational burden can become fairly large, as it depends
heavily on the dimensionality of the parameter space and on the
characteristic of the likelihood function. In certain cosmological
applications, thermodynamic integration can require up to 100
times more likelihood evaluation than parameter
estimation~\citep{Beltran:2005xd}. An elegant algorithm called
``nested sampling'' has been recently put forward by
\cite{SkillingNS}, and implemented in the cosmological context
by~\cite{Bassett:2004wz,Mukherjee:2005wg}. While nested sampling
reduces the number of likelihood evaluations to the same order of
magnitude as for parameter estimation, in the cosmological context
this does not necessarily imply that the computing time can be
reduced accordingly, see~\cite{Mukherjee:2005wg} for details.

\subsection{The Savage--Dickey density ratio}
\label{sec:sddr}

Here we investigate the performance of the Savage-Dickey density
ratio (SDDR), whose use is very promising in terms of reducing the
computational effort needed to calculate the Bayes factor of two
nested models, as we show below (for other possibilities, see
e.g.~\cite{DiCiccio}).

Suppose we wish to compare a two-parameters model $\mdl_1$ with a
restricted submodel $\mdl_0$ with only one free parameter, $\psi$,
and with fixed $\omega = \omega_\star$ (for simplicity of notation
we take a two--parameters case, but the calculations carry over
trivially in the multi--dimensional case). Assume further that the
prior is separable (which is usually the case in cosmology), i.e.\
that
 \begin{equation}
 \pi(\omega, \psi | \mdl_1) = \pi(\omega | \mdl_1) \pi(\psi | \mdl_0).
 \end{equation}
Then the Bayes factor $B_{01}$ of Eq.~\eqref{eq:bfac} can be
written as (see Appendix \ref{app:sddr})
 \begin{equation} \label{eq:savagedickey}
 B_{01} = \left.\frac{\pdf(\omega \vert \data, \mdl_1)}{\pi(\omega |
 \mdl_1)}\right|_{\om = \om_\star} \quad {\text{(SDDR)}}.
 \end{equation}
This expression goes back to J.M.~\cite{dickey71}, who attributed
it to L.J.~Savage, and is therefore called Savage--Dickey density
ratio (SDDR, see also \cite{Verdinelli} and references therein).
Thanks to the SDDR, the evaluation of the Bayes factor of two
nested models only requires the properly normalized value of the
marginal posterior at $\omega = \omega_\star$ under the extended
model $\mdl_1$, which is a by--product of parameter inference. We
note that the derivation of \eqref{eq:savagedickey} {\it does not
involve any assumption about the posterior distribution}, and in
particular about its normality.

For a Gaussian prior centered on $\om_\star$ with standard
deviation $\Delta \om$ and a Gaussian likelihood\footnote{Notice
that $\hat{\mu}$ and $\hat{\sigma}$ are referred to the
likelihood, {\em not} the posterior pdf.} with mean $\hat{\mu}$
and width $\hat{\sigma}$, Eq.~\eqref{eq:savagedickey} gives \be
 \label{eq:B01gauss_text}
\ln B_{01}(\beta, \lambda) = \frac{1}{2}\ln(1 + \beta^{-2}) -
\frac{\lambda^2}{2(1 + \beta^2)},
 \ee
where we have introduced the number of sigma's discrepancy
$\lambda = |\hat{\mu} - \om_\star|/\hat{\sigma}$ and the volume
reduction factor $\beta = \hat{\sigma}/\Delta \om$ (see Appendix
\ref{app:lindley} for details). For strongly informative data,
$\beta^{-1} \gg 1$ and in terms of the information content $I =
-\ln \beta \geq 0$, Eq.~\eqref{eq:B01gauss_text} is approximated
by
 \be \label{eq:B01simple}
 \ln B_{01} \approx I - \lambda^2/2 \quad \text{(informative data)}.
 \ee
The two terms on the right--hand side pull the Bayes factor in
opposite directions: a large information content $I$ signals a
large volume of wasted parameter space under the prior, and acts
as an Occam's razor term favouring the simpler model, while a
large $\lambda$ favours the more complex model because of the
mismatch between the measured and the predicted value of the extra
parameter. Evidence against the simpler model scales as
$\lambda^2$, while evidence in its favour only accumulates as $I =
-\ln \beta$. Furthermore, for strong odds against the simpler
model ($\lambda \gg 1$) the prior choice becomes irrelevant unless
$I \gg \lambda$, a situation which gives rise to Lindley's paradox
(see Appendix \ref{app:lindley}). For the case where $\lambda \ll
1$, \ie~the prediction of the simpler model is confirmed by the
observation, the odds in favour of the simpler model are
determined by the information content $I$, and therefore by the
prior choice.

The use of the SDDR for nested models has several advantages. A
first important point is the analytical insight
Eq.~\eqref{eq:savagedickey} gives into the working of model
selection for two nested models, which we have briefly sketched
above. Priors on the common parameters on both models are
unimportant, as they factor out when computing the Bayes factor.
The only relevant scales in the problem are the quantities
$\lambda$ and $\beta$, see Eq.~\eqref{eq:B01simple}, with the
latter controlled by the prior width on the extra parameter. The
volume effect arising from a change in the prior (e.g., when
enlarging the prior range) can be easily estimated from the SDDR
expression, without recomputing the posterior. Usually, the
posterior pdf in Eq.~\eqref{eq:savagedickey} will be obtained by
Monte Carlo Markov Chain (MCMC)  techniques. In this case, even a
change in the variables, or a more restrictive prior can usually
be applied by simply posterior re--weighting the MCMC samples
without recomputing them. Secondly, the SDDR can be applied to
existing MCMC chains, and therefore the model selection question
can be dealt with easily after the parameter estimation step has
already been performed. Finally, Appendix \ref{app:accuracy}
demonstrates that in the benchmark Gaussian likelihood scenario
the SDDR gives accurate results out to $\lambda \lsim 3$. For
larger value of $\lambda$ the performance of the method is
hindered by the fact that it becomes very difficult with
conventional MCMC methods to obtain samples far out into the tails
of the posterior. One could argue however that the most
interesting regime for model comparison is precisely where the
SDDR can yield accurate answers. This is also the region where
most of the model selection questions in cosmology currently lie.
Finally, often a high numerical accuracy in the Bayes factor does
not seem to be central for most model comparison questions,
especially in view of the fact that the uncertainty in the result
can be strongly dominated by the prior range one assumes. This
suggests that a quick and computationally inexpensive method such
as the SDDR might be helpful in assessing the model comparison
outcome for a broad range of priors. We therefore advocate the use
of SDDR method for model selection questions involving nested
models with moderate discrepancies between the prediction of the
simple model and the posterior result, $\lambda \lsim 3$. We now
turn to the demonstration of the method on current cosmological
observations.

\section{Application to cosmological parameters}
 \label{sec:application}

In this section we apply the Bayesian model selection toolbox
presented above to three cosmological parameters which are central
for our understanding of the cosmological concordance model: the
spectral index of scalar (adiabatic) perturbations, the spatial
curvature of the Universe and an isocurvature cold dark matter
(CDM) component to the initial conditions for cosmological
perturbations.

\subsection{Parameter space and cosmological data}

We use the WMAP 3--year temperature and polarization data
\citep{Hinshaw:2006ia,Page:2006hz} supplemented by small--scale
CMB measurements \citep{Readhead:2004gy,Kuo:2002ua}. We add the
Hubble Space Telescope measurement of the Hubble constant $H_0 =
72 \pm 8$ km/s/Mpc \citep{Freedman:2000cf} and the Sloan Digital
Sky Survey (SDSS) data on the matter power spectrum on linear ($k
< 0.1 h^{-1}\rm Mpc$) scales \citep{Tegmark:2003uf}. Furthermore,
we shall also consider the supernovae luminosity distance
measurements~\citep{Riess:2004nr}. We denote all of the data sets
but WMAP as ``external'' for simplicity of notation. We are also
interested in assessing the changes in the model comparison
outcome in going from WMAP 1--year to WMAP 3--year data. We shall
therefore compare our results using the 3--year WMAP data with the
first--year WMAP data release
\citep{Bennett:2003bz,Hinshaw:2003ex,Verde:2003ey}, complemented
by the ``external'' data sets described above\footnote{A more
detailed discussion on the WMAP first year data model comparison
result and the power of the external data sets can be found in the
original version of the present work,~\cite{Trotta:2005ar}. }.

We make use of the publicly available codes CAMB and CosmoMC
\citep{Lewis:2002ah} to compute the CMB and matter power spectra
and to construct Monte Carlo Markov Chains (MCMC) in parameter
space. The Monte Carlo (MC) is performed using ``normal
parameters'' \citep{Kosowsky:2002zt}, in order to minimize
non--Gaussianity in the posterior pdf. In particular, we sample
uniformly over the physical baryon and cold dark matter (CDM)
densities, $\omega_b \equiv \Omega_b h^2$ and
$\omega_c\equiv\Omega_c h^2$, expressed in units of $1.88\times
10^{-29}~{\rm g/cm}^3$; the ratio of the angular diameter distance
to the sound horizon at decoupling, $\Theta_\star$, the optical
depth to reionization $\tau_r$ (assuming sudden reionization) and
the logarithm of the adiabatic amplitude for the primordial
fluctuations, $\ln 10^{10} A_S$. When combining the matter power
spectrum with CMB data, we marginalize analytically over a bias
$b$ considered as an additional nuisance parameter. Throughout we
assume three massless neutrino families and no massive neutrinos
(for constraints on these quantities, see instead
e.g.~\cite{Bowen:2001in,Spergel:2006hy,Lesgourgues:2006nd}), we
fix the primordial Helium mass fraction to the value predicted by
Big Bang Nucleosynthesis (see e.g.~\cite{Trotta:2003xg}) and we
neglect the contribution of gravitational waves to the CMB power
spectrum.

\subsection{Model selection from current data}

\subsubsection*{The scalar spectral index}

As a first application we consider the scalar spectral index for
adiabatic perturbations, $n_S$. We compare the evidence in favor
of a scale invariant index ($\mdl_0: n_S = 1$), also called an
Harrison-Zel'dovich (HZ) spectrum, with a more general model of
single-field inflation, in which we do not require the spectral
index to be scale invariant, $\mdl_1: n_S \neq 1$. The latter case
is called for brevity ``generic inflation''.

Within the framework of slow--roll inflation, the prior allowed
range for the spectral index can be estimated by considering that
$n_S=1-6 \epsilon+2 \eta$, where $\eta$ and $\epsilon$ are the
slow-roll parameters. If we assume that $\epsilon$ is negligible,
then $n_S = 1 + 2 \eta$. If the slow-roll conditions are to be
fulfilled, $\eta \ll 1$, then we must have $\vert \eta \vert \lsim
0.1$, which gives $0.8 \lsim n_S \lsim 1.2$. Hence we take a
Gaussian prior on $n_S$ with mean $\mu = 1.0$ and width $\sigma =
0.2$.

The result of the model comparison is shown in
Table~\ref{tab:summary}. When employing WMAP 1--year data, the
model comparison yields an inconclusive result ($\ln B_{01} = 0.68
\pm 0.04$), but the new, lower value for $\ns$ from the WMAP
3--year data, enhanced by the small scale CMB measurements and
SDDS matter power spectrum data, does yield moderate evidence for
a non--scale invariant spectral index ($\ln B_{01} = -2.86 \pm
0.28$), with odds of about 17:1, or a posterior probability of a
scale invariant index of 5\%, when compared to the above
alternative generic inflation model. This is a consequence of both
the shift of the peak of the posterior to $\ns = 0.95$ and a
reduction of its spread when using WMAP 3--year data, which places
the scale invariant value of $\ns = 1$ at about $3.3 \sigma$ away
from the posterior's peak (see however the discussion about
possible systematic effects in \cite{Parkinson:2006ku}). In
Table~\ref{tab:summary} we also give the resulting value of the
Bayes factor obtained by using the SDDR formula and a Gaussian
approximation to the posterior, see Eq.~\eqref{eq:B01gauss}. Since
the marginalized posterior for $\ns$ is very well approximated by
a Gaussian, we find a very good agreement between this crude
estimate and the numerical result using the actual shape of the
posterior, with a discrepancy of order $5\%$. This supports the
idea that for reasonably Gaussian pdf's using a Gaussian
approximation to the SDDR might be a good way of obtaining a first
estimate of the Bayes factor for nested models.

Our findings are in broad agreement with~\cite{Parkinson:2006ku},
where it was found using nested sampling that a similar data
compilation as the one employed here gives $\ln B_{01} = -1.99 \pm
0.26$ for the comparison between the HZ model and a generic
inflationary model with a flat prior between $0.8 \leq \ns \leq
1.2$. For such a flat prior, we obtain, using the SDDR, $\ln
B_{01} = -2.98 \pm 0.28$, where the difference
with~\cite{Parkinson:2006ku} has to be ascribed to different
constraining power of the different data compilations used, rather
than to the methods for computing the Bayes factor. For a more
detailed discussion of a series of possible systematic effects
which might change the outcome of the model comparison, see
section IIIC in~\cite{Parkinson:2006ku}.

\begin{table*}
\centering
\begin{minipage}{177mm}
\caption{Summary of model comparison results from WMAP data
combined with small--scale CMB measurements, SDDS, HST and SNIa
data. WMAP3+ext refers to WMAP 3 year data release, WMAP1+ext to
WMAP 1st year data. The most spectacular improvement from WMAP1 to
WMAP3 is the moderate evidence against a scale--invariant spectral
index. Errors in the Bayes factor are obtained by computing the
variance of the SDDR estimate from 5 subchains (see
Appendix~\ref{app:accuracy} for details)\label{tab:summary}. The
``estimate'' column gives the value obtained by employing the
Gaussian approximation to the likelihood, Eq.~\eqref{eq:B01gauss}
for a Gaussian prior or Eq.~\eqref{eq:B01flat} for a flat prior.}
\begin{tabular}{l| c c c c l }
Data  & \multicolumn{2}{c}{$\ln B_{01}$ from SDDR}    & Odds in favour   & Probability & Comment\\
            &(numerical) & (estimate)                 & of simpler model & of simpler model & \\
\hline
 & \multicolumn{5}{c}{Spectral index: $\ns=1$  versus $0.8
\leq \ns \leq 1.2$ (Gaussian)} \\\hline
WMAP3+ext   & $-2.86 \pm 0.28$     & $-3.00$   & 1 to 17  & 0.05 & Moderate evidence for non--scale invariance\\
WMAP1+ext        & $ 0.68 \pm 0.04$     & $0.71$   & 2 to 1 & 0.66
& Inconclusive result\\ \hline
  & \multicolumn{5}{c}{Spatial curvature: $\omk=0$  versus $-1.0
\leq \omk \leq 1$ (Flat)} \\\hline
WMAP3+ext   & $3.37 \pm 0.05$     & $3.25$   & 29 to 1 & 0.97 & Moderate evidence for a flat Universe\\
WMAP1+ext   & $2.70 \pm 0.09$     & $2.68$   & 15 to 1 & 0.94 &
Moderate evidence for a flat Universe\\ \hline
                & \multicolumn{5}{c}{Adiabaticity: $\fiso=0$  versus $-100
\leq \fiso \leq 100$ (Flat)} \\\hline
WMAP3+ext     & $7.62 \pm 0.02$     & $7.63$   & 2050 to 1 &  0.9995 & Strong evidence for adiabatic conditions\\
WMAP1+ext     & $7.50 \pm 0.03$     & $7.53$   & 1800 to 1 &  0.9994 & Strong evidence for adiabatic conditions\\
\end{tabular}
\end{minipage}
\end{table*}

\subsubsection*{The spatial curvature}

We now turn to the issue of the geometry of spatial sections. We
evaluate the Bayes factor for $\Omega_\kappa = 0$ (flat Universe)
against a model with $\Omega_\kappa \neq 0$. As discussed above,
we only need to specify the prior distribution for the parameter
of interest, namely $\Omega_\kappa$. We choose a flat prior of
width $\Delta \Omega_\kappa = 1.0$ on each side of
$\Omega_\kappa=0$, for we know that the universe is not empty
(thus $\Omega_\kappa < 1.0$, setting aside the case of $\Lambda <
0$) nor largely overclosed (therefore $\Omega_\kappa \gsim -1$ is
a reasonable range, see \ref{sec:priors} for further comments).

Cosmic microwave background data alone cannot strongly constrain
$\Omega_\kappa$ because of the fundamental geometrical degeneracy.
Even CMB and SDSS data together allow for a wide range of values
for the curvature parameter, which translates into approximately
equal odds for the curved and flat models. Adding SNIa
observations drastically reduces the range of the posterior, since
their degeneracy direction is almost orthogonal to the geometrical
degeneracy of the CMB. Further inclusion of the HST measurement
for the Hubble parameter narrows down the posterior range
considerably, since the handle on the value of the Hubble constant
today breaks the geometrical degeneracy. When all of the data
(WMAP3+ext) is taken into account, we obtain for the Bayes factor
$\ln B_{01} = 3.37 \pm 0.05$, favouring a flat Universe model with
moderate odds of about $29:1$ (see Table~\ref{tab:summary}). This
corresponds to a posterior probability for a flat Universe of
97\%, for our particular choice of prior. We notice the slight
improvement in these odds from the result obtained using WMAP1+ext
data, where the odds were $15:1$, which is to be ascribed mainly
to the inclusion of polarization data that helps further
tightening constraints around the geometrical degeneracy.

\subsubsection*{The CDM isocurvature mode}

The third case we consider is the possibility of a cold dark
matter (CDM) isocurvature contribution to the primordial
perturbations. For a review of the possible isocurvature modes and
their observational signatures, see e.g.~\cite{Trotta:2004qj}.
Determining the type of initial conditions is a central question
for our understanding of the generation of perturbations, and has
far reaching consequences for the model building of the physical
mechanisms which produced them. Constraints on the isocurvature
fraction have been derived in several works, which considered
different phenomenological mixtures of adiabatic and isocurvature
initial conditions
\citep{Pierpaoli:1999zj,Amendola:2001ni,Trotta:2001yw,
Trotta:2002iz,Trotta:2004,Bucher:2004an,Crotty:2003rz,Valiviita:2003ty,Beltran:2004uv,Moodley:2004nz,Kurki-Suonio:2004mn}.
Two recent studies making use of the latest CMB data
\citep{Bean:2006qz,Keskitalo:2006qv} obtain different conclusions
as to the level of isocurvature contribution. While both groups
report a lower best fit chi--square for a model with a large ($n
\sim 3$) spectral index for the CDM isocurvature component, they
give a different interpretation of the statistical significance of
the improvement. It is precisely in such a context that a model
selection approach as the one presented here might be helpful, in
that it allows to account for the Occam's razor effect described
above. The question of isocurvature modes has been addressed from
a model comparison perspective
by~\cite{Beltran:2005xd,Trotta:2006ww}.

Since the goal of this work is not to present a detailed analysis
of isocurvature contributions, but rather to give a few
illustrative applications of Bayesian model selection, we restrict
our attention to the comparison of a purely adiabatic model
against a model containing a CDM isocurvature mode totally
correlated or anti--correlated. For simplicity, we also take the
isocurvature and adiabatic mode to share the same spectral index,
$n_S$. This phenomenological set--up is close to what one expects
in some realizations of the curvaton scenario, see
e.g.~\cite{Gordon:2002gv,Lyth:2003ip,Lazarides:2004we}. For an
extended treatment including all of the 4 different isocurvature
modes, see~\cite{Trotta:2006ww}.

We compare model $\mdl_0$, with adiabatic fluctuations only, with
$\mdl_1$, which has a totally (anti)correlated isocurvature
fraction
 \be
 \fiso \equiv \frac{ \mathcal{S} }{\zeta},
 \ee
where $\zeta$ is the primordial curvature perturbation and
$\mathcal{S}$ the entropy perturbation in the CDM component (see
\cite{Trotta:2004qj,Lazarides:2004we} for precise definitions).
The sign of the parameter $\fiso$ defines the type of correlation.
We adopt the convention that a positive (negative) correlation,
$\fiso > 0$ ($\fiso < 0$), corresponds to a negative (positive)
value of the adiabatic--isocurvature CMB correlator power spectrum
on large scales. We choose $\fiso$ as the relevant parameter for
model comparison because of its immediate physical interpretation
as an entropy--to--curvature ratio, but this is only one among
several possibilities.

In the absence of a specific model for the generation of the
isocurvature component, there is no cogent physical motivation for
setting the prior on $\fiso$. A generic argument is given by the
requirement that linear perturbation theory be valid, i.e.~$\zeta,
\mathcal{S} \ll 1$. This however does not translate into a prior
on $\fiso$, unless we specify a lower bound for the curvature
perturbation. In general, $\fiso$ is essentially a free parameter,
unless the theory has some built--in mechanism to set a scale for
the entropy amplitude. This however requires digging into the
details of specific realizations for the generation of the
isocurvature component. For instance, the curvaton scenario
predicts a large $\fiso$ if the CDM is produced by curvaton decay
and the curvaton does not dominate the energy density, in which
case $\vert \fiso \vert \sim r^{-1} \gg 1$, since the curvaton
energy density at decay compared with the total energy density is
small, $r \equiv \rho_{\text{curv}}/\rho_{\text{tot}} \ll 1$
\citep{Lyth:2003ip,Gordon:2002gv}. Once the details of the
curvaton decay are formulated, it might be possible to argue for a
theoretical lower bound on $r$, which gives the prior range for
the predicted values of $\fiso$.

In the absence of a compelling theoretical motivation for setting
the prior, we can still appeal to another piece of information
which is available to us before we actually see any data: the
expected sensitivity of the instrument. By assessing the possible
outcomes of a measurement given its forecasted noise levels we can
limit the {\em a priori} accessible parameter space {\em for a
specific observation} on the grounds that it is pointless to admit
values which the experiment will not be able to measure. For the
case of $\fiso$, there is a lower limit to the {\em a priori}
accessible range dictated by the fact that a small isocurvature
contribution is masked by the dominant adiabatic part. Conversely,
the upper range for $\fiso$ is reached when the adiabatic part is
hidden in the prevailing isocurvature mode. In order to quantify
those two bounds, we carry out a Fisher Matrix forecast assuming
noise levels appropriate for the measurement under consideration,
thus determining which regions of parameter space is accessible to
the observation. Such a prior is therefore motivated by the
expected sensitivity of the instrument, rather then by theory. The
prior range for a scale--free parameter thereby becomes a
computable quantity which depends on our prior knowledge of the
experimental apparatus and its noise levels.

\begin{figure}
\centering
\includegraphics[width=\linewidth]{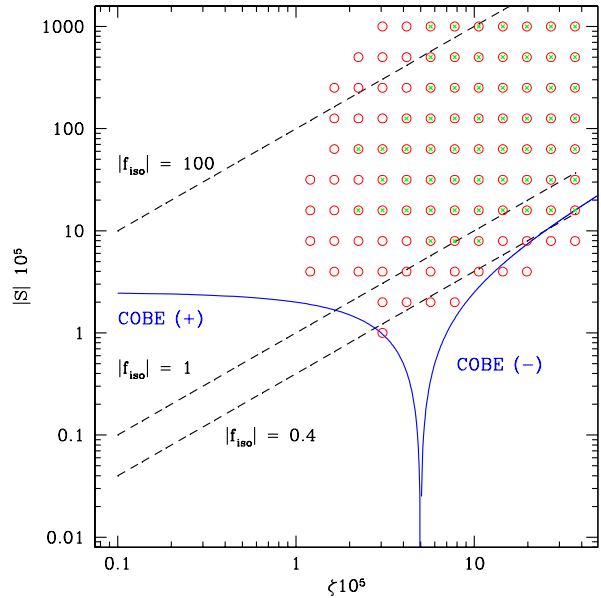}
\caption{The parameter space accessible {\em a priori} to WMAP in
the $(\zeta,|\mathcal{S}|)$ plane is obtained by requiring better
than $10\%$ accuracy on $\Afiso$ in the Fisher Matrix error
forecast (open circles for the best case, crosses for the worst
case, depending on the fiducial values of $\tau_r, n_S$ and on the
sign of the correlation). This translates into a prior accessible
range $0.4 \lsim \Afiso \lsim 100$ (diagonal, dashed lines), but
only if $\zeta, \vert \mathcal{S} \vert \gsim 10^{-5}$. Models
which roughly satisfy the COBE measurement of the large scale CMB
anisotropies ($\delta T / T \approx 10^{-5}$) lie on the
blue/solid line and have positive (negative) correlation left
(right) of the cusp.} \label{fig:prior_WMAP}
\end{figure}

We have performed a FM forecast in the $(\zeta,|\mathcal{S}|)$
plane, whose results are plotted in Figure~\ref{fig:prior_WMAP}
for the WMAP expected sensitivity. We use a grid equally spaced in
the logarithm of the adiabatic and isocurvature amplitudes, in the
range  $10^{-6} \leq \zeta \leq 5\cdot10^{-4}$ and $10^{-8} \leq
|\mathcal{S}| \leq 10^{-2}$. For each pair $(\zeta,|\mathcal{S}|)$
the FM yields the expected error on the amplitudes as well as on
$\fiso$. The expected error however also depends on the fiducial
values assumed for the remaining cosmological parameters. In order
to take this into account, at each point in the
$(\zeta,|\mathcal{S}|)$ grid we run 40 FM forecasts changing the
type of correlation ($\text{sign}(\mathcal{S}) = \pm 1$), the
spectral index ($n_S = 0.8\dots1.2$ with a step of $0.1$) and the
optical depth to reionization ($\tau_r = 0.05\dots0.35$ with a
step of $0.1$). The other parameters ($\theta, \om_c, \om_b$) are
fixed to the concordance model values, since $\zeta, \mathcal{S}$
are mostly correlated with $\tau_r, n_S$ and thus only the
fiducial values assumed for the latter two parameters have a
strong impact on the predicted errors of the amplitudes. We then
select the best and worst outcome for the expected error on
$\fiso$, in order to bracket the expected result of the
measurement independently on the fiducial value for $\tau_r, n_S$.
Notice that at no point we make use of real data. By requiring
that the expected error on $\fiso$ be of order $10\%$ or better,
we obtain the {\em a priori} accessible area in amplitude space
for WMAP, which is shown in Figure~\ref{fig:prior_WMAP}.

It is apparent that $\fiso$ cannot be measured by WMAP if either
$\zeta$ or $\vert S \vert$ are below about $10^{-5}$, in which
case the signal is lost in the detector noise. For amplitudes
larger than $10^{-5}$, $\Afiso = 1$ is accessible to WMAP with
high signal--to--noise independently on the value of $\tau_r,
n_S$, while $\Afiso \approx 0.4$ can be measured only in a few
cases for the most optimistic choice of parameters. As an aside,
we notice that if we restrict our attention to models which
roughly comply with the COBE measurement of the large scale CMB
power (blue/solid lines in Figure~\ref{fig:prior_WMAP}), then WMAP
can only explore the subspace of anti--correlated isocurvature
contribution (right of the cusp) and only if $\zeta \gsim
7\cdot10^{-5}$. On the other end of the range, we can see that
$\Afiso = 100$ is about the largest value accessible to WMAP, at
least for $\zeta \geq 5\cdot10^{-4},|\mathcal{S}| \geq 10^{-2}$.
There is a simple physical reason for the asymmetry of the
accessible range around $\Afiso = 1$: a small isocurvature
contribution can be overshadowed by the adiabatic mode on large
scales due to cosmic variance, but a subdominant adiabatic mode is
still detectable even in the presence of a much larger
isocurvature part, because the first adiabatic peak at $\ell
\approx 200$ sticks out from the rapidly decreasing isocurvature
power at that scale (at least if the spectral tilt is not very
large, as in our case). In conclusion, the values of $\Afiso$
which WMAP can potentially measure with high signal--to--noise are
approximately bracketed by the range $0.4 \leq \Afiso \leq 100$,
assuming that $\zeta \gsim 10^{-5}$. Given the fact that most of
the prior volume lies above $\Afiso = 1$, we can take a flat prior
on $\fiso$ centered around $\fiso = 0$, with a range $-100 \leq
\fiso \leq 100$, or $\Delta \fiso = 100$. As we shall see below,
it is this large range of {\em a priori} possible values compared
with the small posterior volume which heavily penalizes an
isocurvature contribution due to the Occam's razor behavior of the
Bayes factor.

The marginalized posterior on $\fiso$ from WMAP3+ext data gives a
95\% interval $-0.06 \leq \fiso \leq 0.10$, thus yielding only
upper bounds on the CDM isocurvature fraction, in agreement with
previous works using a similar parameterization
(see~\cite{Trotta:2006ww} for details). The spread of the
posterior is of order $0.1$, which lies an order of magnitude
below the level ($\Afiso = 1$) at which an isocurvature signal
would have stand out clearly from the WMAP noise. The Bayes factor
corresponding to the above choice of prior ($-100 \leq \fiso \leq
100$) is given in Table \ref{tab:summary}, and with $\ln B_{01} =
7.62$ it corresponds to a probability of $0.9995$ (or odds of 2050
to 1) for purely adiabatic initial conditions. This is a
consequence of the large volume of wasted parameter space under
the large prior used here, and a fine example of automatic Occam's
razor built into the Bayes factor. We notice that in order to
obtain a model--neutral conclusion (odds of 1:1) one would have to
choose a prior width below $0.1$, \ie~find a mechanism to strongly
limit the available parameter space for the isocurvature
amplitude~\citep{Trotta:2006ww}. In other words, the introduction
of a new scale--free isocurvature amplitude is generically
unwarranted by data, a feature already remarked by
\cite{Lazarides:2004we}.

\begin{figure}
\centering
\includegraphics[width=\linewidth]{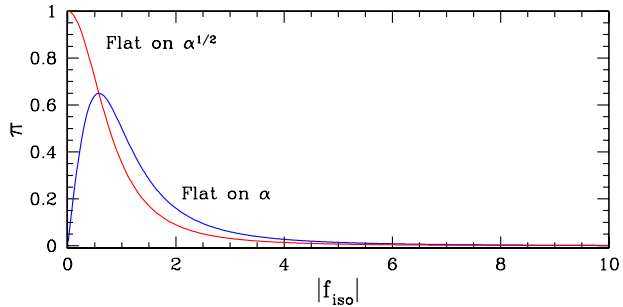}
\caption{Equivalent priors on $\Afiso$ corresponding to the flat
priors used in Beltran et al.~(2005) for the parameters $\alpha$
and $\sqrt{\alpha}$. Both priors cut away the parameter space
$\Afiso \gg 1$, thus reducing the Occam's razor effect caused by a
scale-free parameter. The odds in favor of the purely adiabatic
model thus become correspondingly smaller. Model comparison
results can depend crucially on the variables adopted.}
\label{fig:alphaprior}
\end{figure}

This result differs from the findings of \cite{Beltran:2005xd},
who considered an isocurvature CDM admixture to the adiabatic mode
with arbitrary correlation and spectral tilt and concluded that
there is no strong evidence against mixed models (odds of about
$3:1$ in favor of the purely adiabatic model). While their setup
is not identical to the one presented here and thus a direct
comparison is difficult, we believe that the key reason of the
discrepancy can be traced back to the different basis for the
initial conditions parameter space. Instead of the isocurvature
fraction $\fiso$, \cite{Beltran:2005xd} employ the parameter
$\alpha$ describing the fractional isocurvature power, which is
related to $\fiso$ by
 \be
 \alpha = \frac{\fiso^2}{1+\fiso^2}.
 \ee
The infinite range $0 \leq \vert \fiso \vert < \infty$ corresponds
in this parametrization to a compact interval $[0..1)$ for
$\alpha$ (or $(-1..1)$ for $\sqrt{\alpha}$), over which they take
a flat prior for the variable $\alpha$ (or $\sqrt{\alpha}$). Flat
priors over $\alpha$ or $\sqrt{\alpha}$ correspond to the priors
over $\vert \fiso \vert$ depicted in Figure~\ref{fig:alphaprior},
which cut away the region of parameter space where $\vert \fiso
\vert \gg 1$. As a consequence, the Occam's razor effect is
suppressed and the resulting odds in favor of the purely adiabatic
model are much smaller than in our case.

This example illustrates that model comparison results can depend
crucially on the underlying parameter space. We now turn to
discuss the dependence of our other results on the prior range one
chooses to adopt.

\subsection{Dependence on the choice of prior}
\label{sec:priors}

\begin{figure}
\centering
\includegraphics[width=\linewidth]{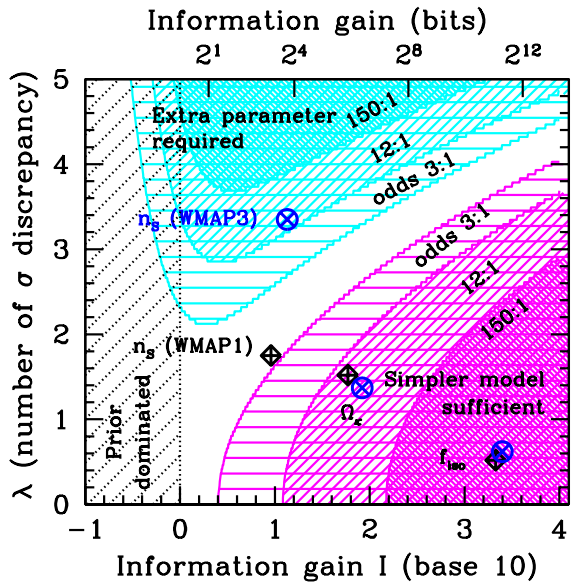}
\caption{Regions in the $(I, \lambda)$ plane (shaded) where one of
the competing models is supported by positive (odds of 3:1),
moderate (12:1) or strong (odds larger than 150:1) evidence. The
white region corresponds to an inconclusive result (odds of about
1:1), while in the region $I<0$ (dotted) the posterior is
dominated by the prior and the measurement is non--informative. In
the lower horizontal axis, $I$ is given in base 10, i.e.~$I = -
\log_{10}\beta$, while it is given in bits in the upper horizontal
axis. The contours are computed from the SDDR formula assuming a
Gaussian likelihood and a Gaussian prior. The location of the
three parameters analyzed in the text is shown by diamonds
(circles) for WMAP1+ext data (WMAP3+ext data). Choosing a wider
(narrower) prior range would shift the points horizontally to the
right (to the left) of the plot.} \label{fig:eplan}
\end{figure}

As described in detail in Appendix \ref{app:lindley}, the Bayes
factor is really a function of two parameters, $\lambda$ and the
information content $I=-\ln \beta$, see Eq.~\eqref{eq:B01gauss}
for the case of a Gaussian prior and a Gaussian likelihood in the
parameter of interest. Figure~\ref{fig:eplan} shows contours of
$\vert \ln B_{01}\vert = \text{const}$ for $\text{const} =
1.0,2.5,5.0$ in the $(I, \lambda)$ plane, as computed from
Eq.~\eqref{eq:B01gauss}. The contours delimit significative levels
for the strength of evidence, as summarized in
Table~\ref{Tab:Jeff}. In the following, we will measure the
information content $I$ in base--10 logarithm. For moderately
informative data ($I \approx 1 - 2$) the measured mean has to lie
at least about $4\sigma$ away from $\om_\star$ in order to
robustly disfavor the simpler model (i.e., $\lambda \gsim 4$).
Conversely, for $\lambda \lsim 3$ highly informative data ($I
\gsim 2$) do favor the conclusion that $\om = \om_\star$. In
general, a large information content favors the simpler model,
because Occam's razor penalizes the large volume of ``wasted''
parameter space of the extended model. A large $\lambda$ disfavors
exponentially the simpler model, in agreement with the sampling
theory result. The location on the plane of the three cases
discussed in the text (the scalar spectral index, the spatial
curvature and the CDM isocurvature component) is marked by
diamonds (circles) for WMAP1+ext (WMAP3+ext). Even though the
informative regions of Figure~\ref{fig:eplan} assume a Gaussian
likelihood, they are illustrative of the results one might obtain
in real cases, and can serve as a rough guide for the Bayes factor
determination.

Another useful properties of displaying the result of the model
comparison in the $(I,\lambda)$ plane as in Figure~\ref{fig:eplan}
is that the impact of a change of prior can be easily estimated. A
different choice of prior will amount to a {\em horizontal shift}
of the points in Figure~\ref{fig:eplan}, at least as long as $I>0$
(\ie, posterior dominated by the likelihood). Thus we can see that
given the results with the priors used in this paper, {\em no
other choice of priors} for $\fiso$ or $\omk$ within 4 order of
magnitude will achieve a reversal of the conclusion regarding the
favoured model. At most, picking more restrictive priors
(reflecting more predictive theoretical models) would make the
points for $\fiso$ or $\omk$ drift to the left of
Figure~\ref{fig:eplan}, eventually entering in the white,
inconclusive region $I \lsim 0.5$. For the spectral index from
WMAP 3--year data, choosing a prior 2 orders of magnitude larger
than the one employed here, ie $-19 < \ns < 20$ would reverse the
conclusion of the model selection, favouring the model $\ns = 1$
with odds of about 3:1. This choice of prior is however physically
unmotivated. On the other hand, reducing the prior by one order of
magnitude -- \ie, making it of the same order as the current
posterior width ($I=0$) -- would still not alter the conclusion
that $\ns = 1$ is disfavoured with moderate odds.

The prior assignment is an irreducible feature of Bayesian model
selection, as it is clear from its presence in the denominator of
Eq.~\eqref{eq:savagedickey}. There is a vast literature which
adresses the problem of assigning prior probabilities (see
footnote \ref{priorrefs}) in a way which reflects the state of
knowledge before seeing the data. In applications to model
selection, it might be more useful to regard the prior as
expressing the available parameter space under the model, rather
then a state of knowledge before seeing the data, as argued in
\cite{Kunz:2006mc}. The underpinnings of the prior choice can be
found in our understanding of model--specific issues. In this work
we have offered two examples of priors stemming from theoretical
motivations: the prior on the scalar spectral index is a
consequence of assuming slow--roll inflation while the prior on
the spatial curvature comes from our knowledge that the Universe
is not empty (and therefore the curvature must be smaller than
$-1$) nor overclosed (or it would have recollapsed). This simple
observations set the correct scale for the prior on
$\Omega_\kappa$, which is of order unity. On the other hand, if
one wanted to impose an inflation--motivated prior of width $\ll
1$, then the information content of the data would go to 0 and the
outcome of the model selection would be non-informative. In
general, it is enough to have an order of magnitude estimate of
the {\em a priori} allowed range for the parameter of interest,
since the logarithm of the model likelihood is proportional to the
logarithm of the prior range. Furthermore, considerations of the
type outlined above can help assessing the impact of a prior
change on the model comparison outcome. Often one will find that
most ``reasonable'' prior choices will lead to qualitatively to
the same conclusion, or else to a non--committal result of the
model comparison.

For essentially scale--free parameters, such as the adiabatic and
isocurvature amplitudes of our third application, model
theoretical considerations of the type employed by
\cite{Lazarides:2004we} can lead to a limitation of the prior
range. In the context of phenomenological model building, we have
demonstrated that an analysis of the {\em a priori} parameter
space accessible to the instrument can be used to define a prior
encapsulating our expectations on the quality of the data we will
be able to gather.

An important {\em caveat} is the dependence of the Bayes factor on
the basis one adopts in parameter space, which sets the natural
measure on the parameters. A flat prior on $\params$ does not
correspond to a flat prior on some other set $\alpha(\params)$
obtained via a non--linear transformation, since the two prior
distributions are related via
\begin{equation}
\pi(\params|\mdl)= \pi({\alpha}|\mdl) \Big\vert
\frac{\dr{\alpha}(\params)} {\dr\params} \Big\vert.
\label{eq:prior}
\end{equation}
As illustrated by the case of the isocurvature amplitude, this is
especially relevant for parameters which can vary over many orders
of magnitudes. We put forward that the choice of the parameter
basis can be guided by our physical insight of the model under
scrutiny and our understanding of the observations. This principle
would suggest that one should adopt flat priors along ``normal
variables'' or principal components, because those are directly
probed by the data and usually can be interpreted in terms of
physically relevant and meaningful quantities. A general principle
of consistency can be invoked to select the most appropriate
variable for cases where many apparently equivalent choices are
present (for example, $\fiso$, $\alpha$ or $\sqrt{\alpha}$). We
leave further exploration of this very relevant issue to a future
publication.

\section{Conclusions}
\label{sec:conclusions}

We have argued that frequentist significance tests should be
interpreted carefully and in particular that Bayesian model
selection reasoning should be used to decide whether the
introduction of a parameter is warranted by data. The main
strengths of the Bayesian approach are that it does consider the
information content of the data and that it allows one to confirm
predictions of a model, instead of just disproving them as in the
sampling theory approach.

We have investigated the use of the Savage--Dickey density ratio
(SDDR) as a tool to compute the Bayes factor of two nested models,
at no extra computational cost than the Monte Carlo sampling of
the parameter space. The technique is likely to be accurate for
cases where the the estimated value of the extra parameter under
the larger model lies less than about 3 sigma's away from the
predicted value under the simpler model, as shown in
Appendix~\ref{app:accuracy}. In a companion paper
\citep{Trottaprep} a complementary technique is introduced, called
PPOD, which produces forecasts for the probability distribution of
the Bayes factor from future experiments.

We have applied this Bayesian model selection point of view to
three central ingredients of present--day cosmological model
building. Regarding the spectral index of scalar perturbations, we
found that WMAP 3--year data disfavour a scale--invariant spectral
index with moderate evidence, and that this result holds true for
all reasonable choices of priors. This is a significant change
with respect to the inconclusive result one obtained using the
WMAP 1st year data release instead. We found that the odds in
favour of a flat Universe have doubled (from $15:1$ to $29:1$) in
going from WMAP1+ext to WMAP3+ext, and we have stressed that this
conclusion can only be obtained if the Hubble parameter is
measured independently or if supernovae luminosity distance
measurements (or other low--redshift rulers, such as baryonic
acoustic oscillations, see~\cite{Eisenstein:2005su}) are employed.
Finally, purely adiabatic initial conditions are strongly
preferred to a mixed model containing a totally (anti)correlated
CDM isocurvature contribution (odds larger than $1000:1$), on the
grounds of an Occam's razor argument, that the prior available
parameter space is much larger than the small surviving posterior
volume. This is however crucially dependent on the variable one
chooses to impose flat priors on.

In the light of these findings, it seems to us that model
comparison tools offer complementary insight in what the data can
tell us about the plausibility of theoretical speculations
regarding cosmological parameters, and can provide useful guidance
in the quest of a cosmological concordance model.

{\em Acknowledgments} I am grateful to Chiara Caprini, Ruth
Durrer, Samuel Leach, Julien Lesgourgues  and Christophe Ringeval
for useful discussions. I thank  Martin Kunz, Andrew Liddle, Tom
Loredo, Pia Mukherjee and David Parkison for many enlightening
discussions and valuable comments on earlier drafts. I thank an
anonymous referee for many helpful suggestions. This research is
supported by the Tomalla Foundation, by the Royal Astronomical
Society through the Sir Norman Lockyer Fellowship and by St Anne's
College, Oxford. The use of the Myrinet cluster (University of
Geneva) and of the Glamdring cluster (Oxford University) is
acknowledged. I acknowledge the use of the package
\texttt{cosmomc}, available from \texttt{cosmologist.info},  and
the use of the Legacy Archive for Microwave Background Data
Analysis (LAMBDA). Support for LAMBDA is provided by the NASA
Office of Space Science.


\appendix

\section{An illustration of Lindley's paradox}
\label{app:lindley}

\cite{Lindley}'s paradox describes a situation where frequentist
significance tests and Bayesian model selection procedures give
contradictory results. As we demonstrate below, it arises because
the information content of the data is neglected in the
frequentist approach.

Let us consider the toy example of a measurement of a Gaussian
distributed quantity, $\omega$, by drawing $n$ iid samples with
known s.d. $\sigma$. Then the likelihood function is the normal
distribution
 \be
 p(\hat{\mu}, \hat{\sigma} | \omega) = \norm{\hat{\mu}}{\hat{\sigma}}{\om},
 \ee
where $\hat{\mu}$ is the estimated mean and
$\hat{\sigma}=\sigma/\sqrt{n}$ its uncertainty. From the point of
view of frequentist statistics, a significance test is performed
on the null hypothesis $\mathcal{H}_0: \omega = \om_\star$. We
define $\lambda$ as dimensionless number which indicates ``how
many sigma's away'' is our estimate of the mean, $\hat{\mu}$, from
its value under $\mathcal{H}_0$ in units of its uncertainty:
 \begin{equation}
 \label{eq:lambda}
 \lambda \equiv \frac{|\hat{\mu} - \om_\star |}{\hat{\sigma}}.
\end{equation}
This ``number of sigma's'' difference is interpreted as a measure
of the confidence with which one can reject $\mathcal{H}_0$. The
``p--value'' \be \label{eq:pvalue}
 \text{p--value} = \int_{\lambda}^\infty p(\hat{\mu}, \hat{\sigma} | \omega)\dr
 \om
 \ee
is compared to a number $\alpha$, called the ``significance
level'' of the test and the hypothesis $\mathcal{H}_0$ is rejected
at the $1-\alpha$ confidence level if $\text{p--value} < \alpha$.
If we pick a (fixed) confidence level, say $\alpha = 0.05$, then
the frequentist significance test rejects the null hypothesis if
 \be \label{eq:sigtes}
 Z(\lambda) \equiv \frac{1}{\sqrt{2\pi}} \int_\lambda^\infty \exp\left(
 -t^2/2\right) \dr t  \leq \alpha/2 .
 \ee
(for a 2--tailed test). For $\alpha = 0.05$ the equality in Eq.
\eqref{eq:sigtes} holds for $\lambda = 1.96$. In other words,
sampling statistics reject the null hypothesis at the 95\%
confidence level if the measured mean is more than $\lambda =
1.96$ sigma's away from the predicted $\om_\star$ under
$\mathcal{H}_0$.

This conclusion can be in strong disagreement with the Bayesian
evaluation of the Bayes factor, i.e. a value $\om_\star$ rejected
under a frequentist test can on the contrary be favored by
Bayesian model comparison~\citep{Lindley}. In the Bayesian model
comparison approach, the two competing models are $\mdl_0$, with
no free parameters, in which the value of $\omega$ is fixed to
$\omega = \om_\star$, and model $\mdl_1$, with one free parameter
$\omega \neq \om_\star$. Under $\mdl_1$, our prior belief before
seeing the data on the probability distribution of $\omega$ is
explicitly represented by the prior pdf $\pi(\omega)$. This prior
pdf is then updated to the posterior via Bayes
theorem\footnote{Notice that, after applying Bayes theorem, the
posterior probability is attached to the parameter $\omega$
itself, not to the estimator $\hat{\mu}$ as in sampling theory. In
the Bayesian framework we only deal with observed data, never with
properties of estimators based on a (fictional) infinite
replication of the data. In cosmology one only has one realization
of the Universe and there is not even the conceptual possibility
of reproducing the data {\em ad infinitum} and therefore the
Bayesian standpoint seems better suited to such a situation.},
Eq.~\eqref{eq:Bayes_Theorem}.

A formal measure of the information gain obtained through the data
is the cross--entropy between prior and posterior, the
Kullback--Leibler divergence
 \be
 \KL(p,\pi) = \int \pdf(\params \vert \data)
 \ln \frac{\pdf(\params \vert \data)}{\pi(\params)} \dr \params  .
 \ee
For a Gaussian prior of standard deviation $\Delta \om$ centered
on $\om_\star$ and a Gaussian likelihood with mean $\hat{\mu}$ and
standard deviation $\hat{\sigma}$, the information gain is given
by
 \be \label{eq:KL}
 \KL + \frac{1}{2} = - \ln \beta + \frac{1}{2}\beta^2
 (\lambda^2-1),
 \ee
where we have defined
 \be
 \beta \equiv \hat{\si}/\Delta \om,
 \ee
the factor by which the accessible parameter space under $\mdl_1$
is reduced after the arrival of the data (remember that
$\hat{\sigma}$ is the standard deviation of the likelihood). For
totally uninformative data, $\beta = 1$ and $\lambda = 0$, and
thus $\KL = 0$. Unless $\lambda \gg 1$ (in which case the null
hypothesis is rejected with many sigma's and there is hardly any
need for model comparison) we can usually neglect the second term
on the right--hand--side of Eq.~\eqref{eq:KL}. We are therefore
led to define a simpler measure of the {\em information content}
of the data, $I$, as
\begin{equation} \label{eq:infocontent}
I \equiv -\ln \beta.
\end{equation}
The choice of the logarithm base is only a matter of convenience,
and this sets the units in which the entropy is measured. Had we
chosen base--2 logarithm instead, the information would have been
measured in bits. In Figure~\ref{fig:eplan}, the choice of using
the base--10 logarithm for the bottom horizontal axis means that
$I$ describes the order of magnitude by which our prior knowledge
has improved after the arrival of the data.

We now compute the Bayes factor $B_{01}$ in favor of model
$\mdl_0$ from Eq.~\eqref{eq:savagedickey}, using again the above
Gaussian prior, obtaining
 \begin{align}
 \label{eq:B01gauss}
 B_{01}(\beta, \lambda) = \sqrt{1 + \beta^{-2}} \exp \left(
 - \frac{\lambda^2}{2(1 + \beta^2)}\right).
 \end{align}
The model comparison result thus depends not only on $\lambda$,
but also on the quantity $\beta$, which is proportional to the
volume occupied by the posterior in parameter space and describes
the information gain in going from prior to posterior. If instead
of a Gaussian prior one takes a flat prior around $\om_\star$ of
width $2\Delta \omega$ (the factor of 2 being chosen to facilitate
the comparison with the case of a Gaussian prior of standard
deviation $\Delta \omega$) one obtains instead
 \begin{align} \label{eq:B01flat}
  B_{01}(\beta, \lambda) & = \sqrt{\frac{2}{\pi}}\beta^{-1}\exp\left(-\lambda^2/2 \right)
  \times \\
  & \left[ Z \left(\lambda - \beta^{-1}\right)
  - Z\left(\lambda + \beta^{-1}\right)
  \right]^{-1}. \notag
  \end{align}
where the function $Z(y)$ is defined in \eqref{eq:sigtes}, a
consequence of the top--hat prior. For $\beta^{-1} \gg \lambda$,
the posterior is well localized within the boundaries of the prior
and the term in square brackets in \eqref{eq:B01flat} tends to
$1$.

\begin{figure}
 \centering
 \includegraphics[width=\linewidth]{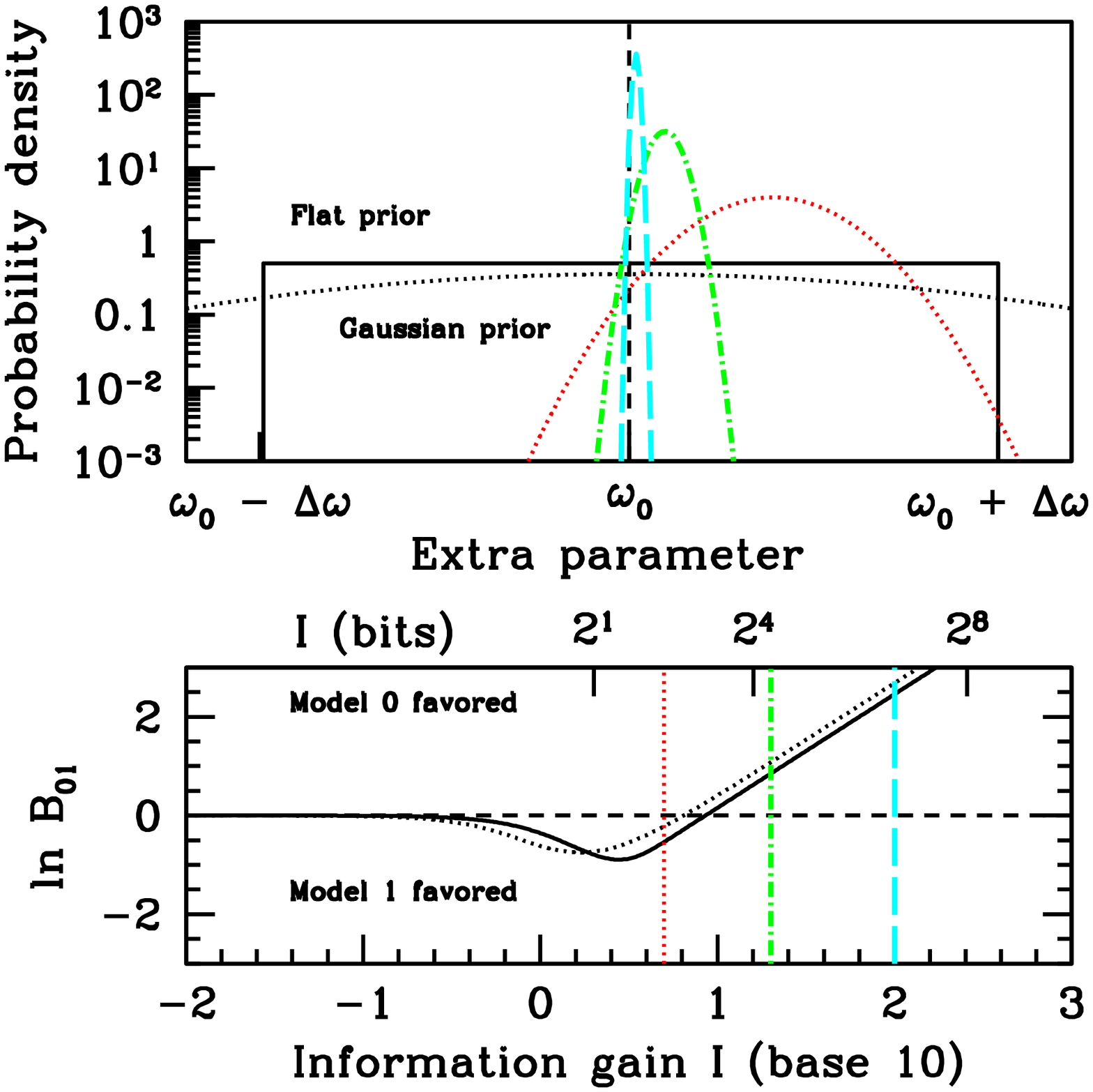}
\caption{Illustration of Lindley's paradox. Sampling statistics
hypothesis testing rejects the hypothesis that $\omega =
\om_\star$ with 95\% confidence in all 3 cases (coloured curves)
illustrated in the top panel ($\lambda = 1.96$ in all cases).
Bayesian model selection does take into account the information
content of the data $I$, and correctly favors the simpler model
(predicting that $\omega = \om_\star$) for informative data (right
vertical line in the bottom panel, $I = 2$ expressed in base--10
logarithm), with odds of $14:1$ (for a Gaussain prior, dotted
black line). Using a flat prior of the same width (solid black
line) instead reduces $\ln B_{01}$ by a geometric factor
$\ln(2/\pi)/2 = 0.22$ in the informative ($I \gg 1$) regime.
Notice that for non--informative data ($I \ll 0$) the Bayes factor
reverts to equal odds for the two
models.}\label{fig:B01illustration}
 \end{figure}

In order to clarify the role of the information content and the
difference with frequentist hypothesis testing, consider the
following example (see Figure \ref{fig:B01illustration}). For a
fixed choice of prior width $\Delta \omega$, imagine performing
three different measurements, each with a different value of
$\beta$ (i.e., with different information content $I$) but with
outcomes such that $\lambda$ is the same in all three cases. This
is depicted in the top panel of Figure \ref{fig:B01illustration},
where the likelihood mean is $\lambda = 1.96$ sigma's away from
$\om_\star$ for all three cases. Under sampling statistics, all
three measurements equally reject the null hypothesis, that
$\omega=\om_\star$, at the 95\% confidence level. And yet common
sense clearly tells us that this cannot be the right conclusion in
all three cases. Indeed, the Bayes factor, Eq.~\eqref{eq:B01gauss}
or \eqref{eq:B01flat}, correctly recovers the intuitive result
(bottom panel of Figure \ref{fig:B01illustration}): the
measurement with the larger error ($\beta=1/5$, or $I=0.7$,
expressed in base--10 logarithm) corresponds to the least
informative data, and the Bayes factor slightly disfavours the
simpler model ($\ln B_{01} = -0.2$, or odds of about 5:4 against
$\mdl_0$ and $p(\mdl_0|\data) = 0.44$). For $\beta = 1/20$ or
$I=1.3$ (moderately informative data), evidence starts to
accumulate {\em in favor} of $\mdl_0$ ($\ln B_{01} = 1.08$, odds
of 3:1 in favor and $p(\mdl_0 |\data) = 0.75$). For very
informative data, $\beta = 1/100, I= 2$, Bayesian reasoning
correctly deduces that the simpler $\mdl_0$ should be favored
($\ln B_{01} = 2.68$, odds of 14:1 in favor of $\mdl_0$ and a
posterior probability $p(\mdl_0|\data) = 0.94$). The above numbers
are for a Gaussian prior, but those conclusion are largely
independent of the choice of a Gaussian or of a flat prior,
provided the bulk of the prior volume is the same (compare the
dotted and solid line in the bottom panel of
Figure~\ref{fig:B01illustration} for a Gaussian and a flat prior,
respectively).

This illustration shows that the Bayes factor can correctly favor
models which would be rejected with high confidence by hypothesis
testing in a sampling theory approach. While in sampling theory
one is only able to disprove models by rejecting hypothesis, it is
important to highlight that the Bayesian evidence can and does
accumulate {\em in favor} of simpler models, scaling as $1/\beta$.
While it is easier to disprove $\om = \om_\star$, since model
rejection is exponential with $\lambda$, the Bayesian approach
allows to evaluate what the data have to say {\em in favor} of an
hypothesis, as well.

In summary, quoting the number of sigma's away from $\om_\star$
(the $\lambda$ parameter) is not always an informative statement
to decide whether or not a parameter $\omega$ differs from
$\om_\star$. Answering this question is a model comparison issue,
which requires the evaluation of the Bayes factor.

\section{Derivation of the SDDR}
\label{app:sddr}

The Bayes factor $B_{01}$ of Eq.~\eqref{eq:evidence_def} can be
evaluated by computing the integrals
 \begin{align}
 \pdf(\mdl_0 \vert \data) & = \int \dr \psi \pi_0(\psi) p(\data \vert\psi,
 \om_\star) ,\\
 \pdf(\mdl_1 \vert \data) & = \int \dr \psi \dr \omega \pi_1(\psi, \omega) p(\data \vert\psi,
 \omega) \equiv q.
 \end{align}
Here $\pi_0(\psi)$ denotes the prior over $\psi$ in model
$\mdl_0$, and $\pi_1(\psi, \omega)$ the prior over $(\psi,
\omega)$ under model $\mdl_1$. Note that, since the models are
nested, the likelihood function for $\mdl_0$ is just a slice at
constant $\omega = \om_\star$ of the likelihood function in model
$\mdl_1$, $p(\data \vert\psi, \omega)$.

Now multiply and divide $B_{01}$ by the number $\pdf(\om_\star
\vert \data) \equiv \pdf(\omega = \om_\star \vert \data, \mdl_1)$,
which is the marginalized posterior for $\omega$ under $\mdl_1$
evaluated at $\om_\star$, and using that $\pdf(\om_\star \vert
\data) = \pdf(\om_\star, \psi \vert \data)/\pdf(\psi \vert
\om_\star, \data)$ at all points $\psi$, we obtain
 \begin{align}
 B_{01} & = \pdf(\om_\star \vert \data) \int \dr \psi \frac{\pi_0(\psi) p(\data \vert\psi,
 \om_\star)\pdf(\psi\vert\om_\star, \data)}{q \cdot \pdf(\om_\star, \psi \vert
 \data)}\\
 & = \pdf(\om_\star \vert \data) \int \dr \psi \frac{\pi_0(\psi)
 \pdf(\psi\vert\om_\star, \data)}{\pi_1(\om_\star, \psi)}, \label{eq:intermediate_step}
 \end{align}
where in the second equality we have used the definition of
posterior, namely that $\pdf(\om_\star, \psi \vert \data) =
p(\data \vert\om_\star, \psi)\pi_1(\om_\star, \psi) / q$. Up to
this point we have not made any assumption nor approximation. We
now assume that the prior satisfies
 \begin{equation} \label{eq:prior_assumption}
 \pi_1(\psi \vert \om_\star) = \pi_0(\psi),
 \end{equation}
which always holds in the (usual in cosmology) case of separable
priors, i.e.
 \begin{equation}
 \pi_1(\omega, \psi) = \pi_1(\omega) \pi_0(\psi).
 \end{equation}
Under this assumption, and since $\pdf(\psi\vert\om_\star, \data)$
in \eqref{eq:intermediate_step} is the normalized marginal
posterior, Eq.~\eqref{eq:intermediate_step} simplifies to the SDDR
given in Eq.~\eqref{eq:savagedickey}.

\section{Benchmark tests for the SDDR}
\label{app:accuracy}

In order to explore the accuracy of the SDDR, we have tested its
performance for the benchmark case of a Gaussian likelihood. A
$D$--dimensional likelihood is generated by choosing a random
$D$--dimensional, diagonal covariance matrix. The correlations can
be set to 0 without loss of generality since in the Gaussian case
it is always possible to rotate to the principal axis of the
covariance ellipse. The mean of the likelihood is set to 0 for the
last $D-1$ dimensions, while for the first parameter (the one we
are interested in testing) the mean is chosen to lie
$\lambda\sigma_1$ away from 0, where $\lambda$ is selected below
and $\sigma_1^2$ is the covariance along direction 1. We then
compare the two following nested models: $\mdl_0$ predicts that
the first parameter $\theta_1=0$, while $\mdl_1$ has a Gaussian
prior centered around 0 and of width $\Delta w = \sigma_1/\beta$,
where $\beta$ is fixed.

\begin{figure}
\centering
\includegraphics[width=\linewidth]{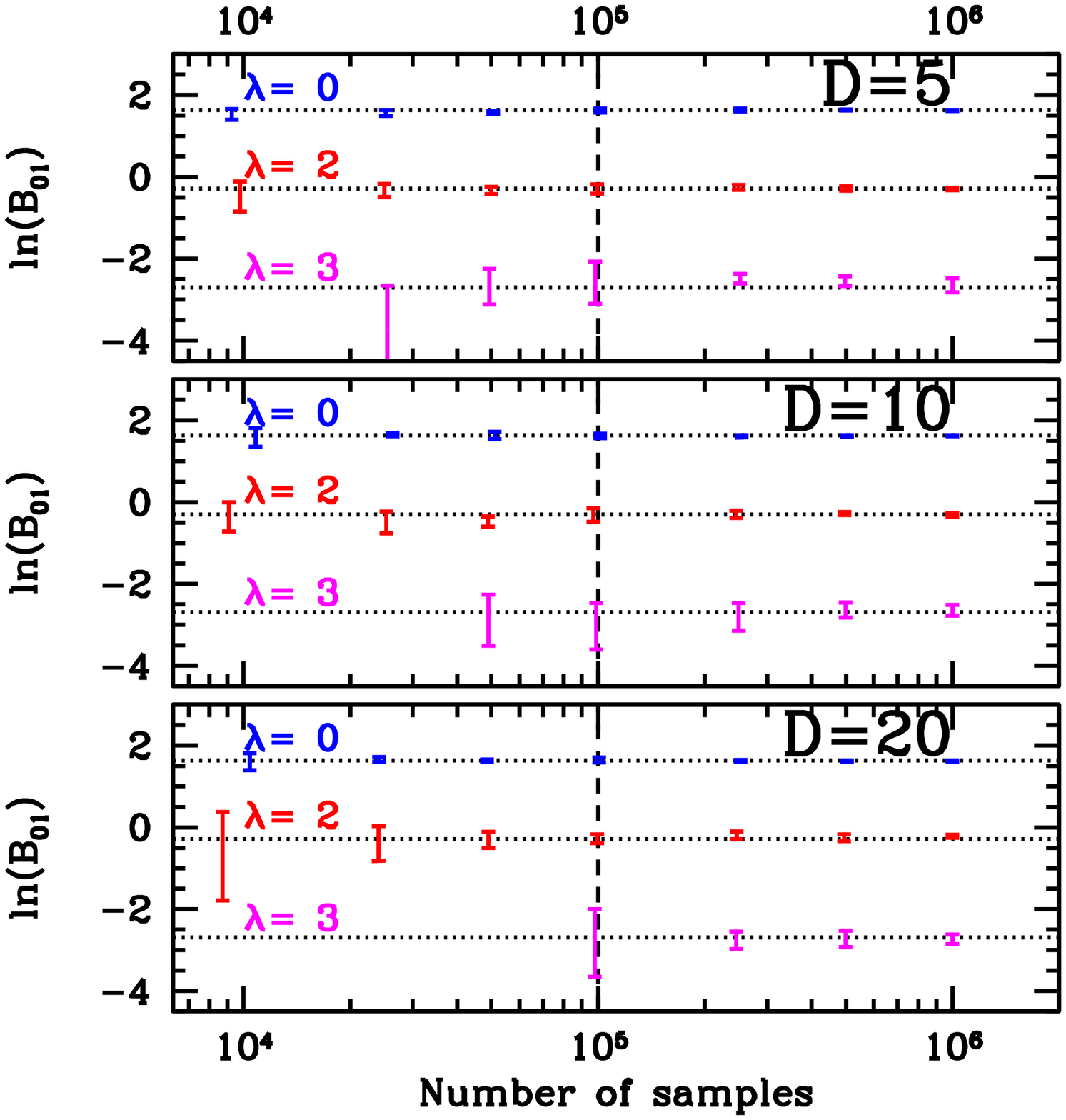}
\caption{Benchmark test for the SDDR formula for a Gaussian
likelihood and prior, for parameter spaces of dimensionality $D$.
The horizontal, dotted lines give the exact value. The SDDR
performs extremely well for comparing models lying $\lambda < 3$
sigma's away from each other. In this case, less than $10^5$
samples are required to achieve a satisfactory agreement with the
exact result. For $\lambda \gsim 4$ the tails of the likelihood
are not sufficiently explored to apply the SDDR. The missing
points for $\lambda = 3$ indicate that the given number of samples
are insufficient to achieve coverage of the simpler model
prediction.} \label{fig:benchmark}
\end{figure}

The posterior is then reconstructed using a MCMC algorithm and the
Bayes factor computed using the SDDR. The results are shown in
Figure~\ref{fig:benchmark} as  a function of the number of samples
for parameter spaces of dimension $D=5,10,20$ and for $\lambda =
1,2,3$. We have fixed $\beta = 0.2$ throughout (changing the value
of $\beta$ only rescales the Bayes factor without affecting the
accuracy, as long as $\beta < 1$, \ie~for informative data). The
errors on the Bayes factor are computed as in the text using a
bootstrap technique: the full sample set is divided in $R=5$
subsets, then the mean and standard deviation of the SDDR are
computed from those subsets. The error thus only reflect the
statistical noise within the chain and it does not take into
account a possible systematic under--exploration of the
likelihood's tails.

It is clear that the SDDR performs extremely well for $\lambda
\leq 2$ while it becomes less accurate for $\lambda = 3$. This is
because it is rather difficult to explore regions further out in
the tails of the distribution using conventional MCMC methods. For
$\lambda > 3$ it becomes very unpractical to obtain sufficient
samples in the tail. For models that lie less than about 3 sigma's
away from each other, the SDDR gives a satisfactory accuracy in
the model comparison result at no extra cost than the parameter
estimation step, requiring less than $10^5$ samples. Furthermore,
the scaling with the dimensionality of the parameter space appears
to be rather favourable, and the error increases only mildly from
$D=5$ to $D=20$ at a given number of samples.

Clearly, for likelihoods that are close to Gaussian, the
approximations \eqref{eq:B01gauss} and \eqref{eq:B01flat} can
still give a useful order of magnitude estimate of the result.
Finally, we stress that in the regime where the SDDR works well
($\lambda \lsim 3$) its accuracy is not limited by the assumption
of normality of the likelihood, but only by the efficiency and
accuracy of the MCMC reconstruction of the posterior. Particular
care must be exercised in exploring accurately distributions
presenting heavier tails than Gaussians, and further work is
required to extend the MCMC sampling to the regime $\lambda \gsim
4$. In this case, sampling at a higher temperature could help in
obtaining sufficient samples in the tail, an issue whose
exploration we leave for future work.

\end{document}